\documentclass[reprint,twocolumn,superscriptaddress,secnumarabic,amssymb, nobibnotes, aps,pra,longbibliography]{revtex4-1}

\usepackage{amsmath}    
\usepackage{graphicx}    
\usepackage{verbatim}    
\usepackage{color}           
\usepackage{xcolor}
\usepackage{subfigure}   
\usepackage{hyperref}    
\usepackage{verbatim}  
\definecolor{ao}{rgb}{0.0, 0.5, 0.0}
\usepackage[normalem]{ulem}

\newcommand{\red}[1]{\textcolor{red}{#1}}
\newcommand{\blue}[1]{\textcolor{blue}{#1}}

\begin{document}


\input epsf

\title{Charge-Transfer Electronic Structure of NiX$_2$ (X = S, Se)}

\author{Atsushi Hariki}
\affiliation{Department of Physics and Electronics, Graduate School of Engineering, Osaka Metropolitan University, 1-1 Gakuen-cho, Nakaku, Sakai, Osaka 599-8531, Japan}

\author{Takaki Okauchi}
\affiliation{Department of Physics and Electronics, Graduate School of Engineering, Osaka Metropolitan University, 1-1 Gakuen-cho, Nakaku, Sakai, Osaka 599-8531, Japan}

\author{Daisuke Takegami}
\affiliation{Department of Physics, Tokyo Metropolitan University, Hachioji, 192-0397, Japan}
\affiliation{Max Planck Institute for Chemical Physics of Solids, N{\"o}thnitzer Str. 40, 01187 Dresden, Germany}

\author{Naoki Ito}
\affiliation{Department of Physics and Electronics, Graduate School of Engineering, Osaka Metropolitan University, 1-1 Gakuen-cho, Nakaku, Sakai, Osaka 599-8531, Japan}

\author{Mizuki Furo}
\affiliation{Department of Physics and Electronics, Graduate School of Engineering, Osaka Metropolitan University, 1-1 Gakuen-cho, Nakaku, Sakai, Osaka 599-8531, Japan}

\author{Ayako~Yamamoto}
\affiliation{College of Engineering, Shibaura Instutite of Technology, Fukasaku 307, Mimuma, Saitama, 337-8570, Japan}
\author{Tomoya~Higo}
\affiliation{Department of Physics, University of Tokyo, 7-3-1 Hongo, Bunkyo, Tokyo 113-0033, Japan}
\author{Satoru~Nakatsuji}
\affiliation{Department of Physics, University of Tokyo, 7-3-1 Hongo, Bunkyo, Tokyo 113-0033, Japan}
\affiliation{Institute for Solid State Physics, University of Tokyo, 5-1-5 Kashiwanoha, Kashiwa, Chiba 277-8581, Japan}

\author{Masato Yoshimura}  
\affiliation{National Synchrotron Radiation Research Center, 101 Hsin-Ann Road, 30076 Hsinchu, Taiwan}

\author{Takashi Mizokawa}  
\affiliation{Department of Applied Physics, Waseda University, 3-4-1 Okubo, Shinjuku-ku, Tokyo 169-8555, Japan}

\author{Jan Kune\v{s}}
\affiliation{Department of Condensed Matter Physics, Faculty of Science, Masaryk University, Kotl\'a\v{r}sk\'a 2, 611 37 Brno, Czechia}

\date{\today}

\begin{abstract}
We investigate the electronic structures of NiS$_2$ and NiSe$_2$ using density functional theory combined with dynamical mean-field theory (DFT+DMFT). 
A realistic electronic structure within DFT+DMFT was determined by optimizing the double-counting correction to reproduce experimental valence-band photoemission spectra. 
The validity of the present model is further confirmed by its successful description of the Ni 2$p$ core-level photoemission and Ni $L$-edge x-ray absorption spectra of NiS$_2$.
Our results reveal a smaller charge-transfer energy than previously assumed, resulting in substantial ligand-to-Ni charge transfer and a reduced Ni local moment.
We clarify how the relative position and interaction between the Ni upper Hubbard band and the antibonding chalcogen-dimer states shape the evolution of the low-energy electronic structure across the NiS$_{2-x}$Se$_x$ series.
\end{abstract}

\maketitle
\section{Introduction}

Competition between intra-atomic electron correlations and chemical bonding mediated by interatomic hybridization is at the heart of many emergent phenomena in correlated materials, including magnetism, metal-insulator transitions (MITs), and superconductivity. In systems with partially filled transition-metal (TM) 3$d$ shells, strong Coulomb interactions can localize electrons and open a gap, producing a Mott insulating state characterized by lower and upper Hubbard bands (LHB and UHB). When ligand degrees of freedom become important, however, the electronic gap may instead be controlled by charge excitations between ligand-derived states and TM $d$ states, leading to a charge-transfer (CT) insulating state within the Zaanen-Sawatzky-Allen (ZSA) classification~\cite{Zaanen1985}.

Nickel dichalcogenides in the pyrite family provide a particularly rich platform for exploring the interplay between electronic correlations, covalent bonding, and structural tuning. NiS$_2$ is commonly classified as a CT insulator, with a small insulating gap of approximately 0.2~eV in the bulk~\cite{Perucchi2009,Sato1984,Kunes2010}. Remarkably, the insulating state survives into the paramagnetic regime, highlighting the essential role of electron correlations beyond conventional band theory. In contrast, the isostructural compound NiSe$_2$ is metallic, and the solid solution NiS$_{2-x}$Se$_x$ exhibits a bandwidth-controlled MIT driven by Se substitution, pressure, or temperature~\cite{Kwizera1980,Yao1996,Matsuura2000,Miyasaka2000,Takeshita2007,Niklowitz2008,Feng2011,Xu2014}. Thus, NiS$_{2-x}$Se$_x$ provides a unique materials platform in which electronic correlations can be tuned continuously without changing the underlying crystal structure, enabling systematic investigations of the mechanisms governing correlation-driven MITs.

\begin{figure}[h]
     \includegraphics[width=0.80\columnwidth]{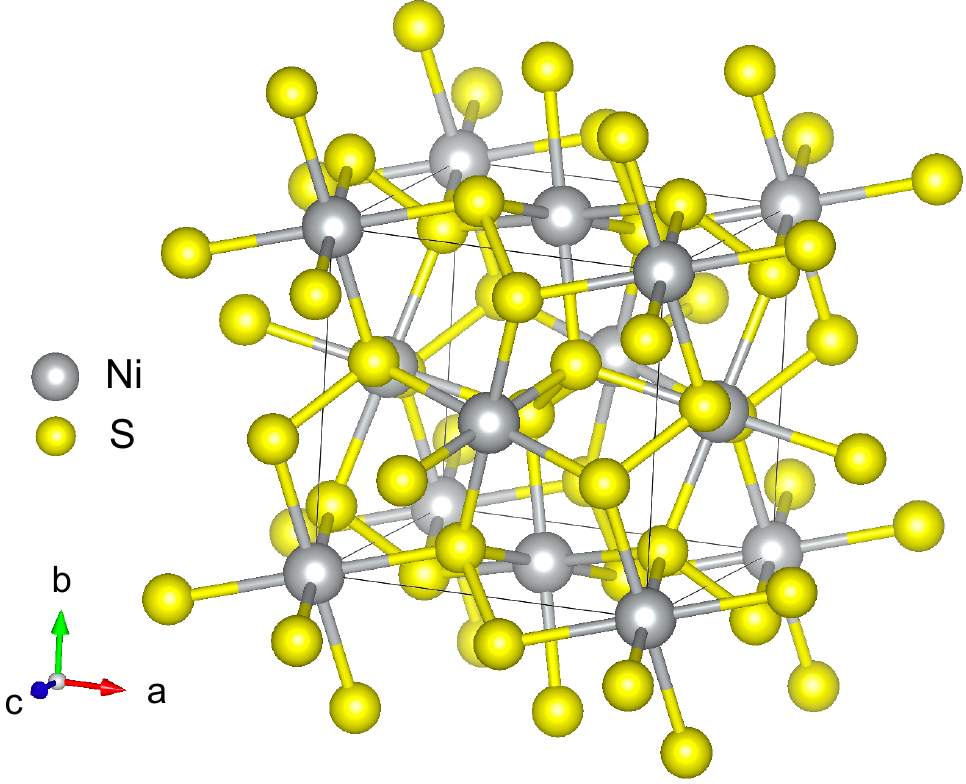}
    \caption{Crystal structure of NiS$_2$. The crystal structure is visualized using VESTA3~\cite{vesta}.}
    \label{fig:struct}
\end{figure}

Both NiS$_2$ and NiSe$_2$ crystallize in the cubic pyrite structure (Fig.~\ref{fig:struct}), consisting of a face-centered cubic lattice of Ni atoms and chalcogen dimers (S$_2$ or Se$_2$). The molecular character of these dimers plays a crucial role in determining the electronic structure. The chalcogen $p_{\sigma}$ orbitals oriented along the dimer axis form bonding and antibonding combinations with a large energy separation, pushing the antibonding states just above the Fermi level. This bonding picture differs from conventional CT oxides such as NiO, where occupied O 2$p$ states hybridize with Ni 3$d$ orbitals and the conduction-band edge is primarily associated with the UHB. The strong sensitivity of the chalcogen dimer states to chemical substitution and lattice compression makes the pyrite NiS$_{2-x}$Se$_x$ ideal systems for studying the interplay between covalency, bandwidth control, and correlation effects.

Recently, the nickel dichalcogenides were investigated in the context of Hund's metal physics~\cite{Jang2021,Park2024} and as a potential altermagnetic platform~\cite{park2026}. Several studies using density functional theory (DFT) and its combination with dynamical mean-field theory (DFT+DMFT) have been performed for the NiS$_{2-x}$Se$_x$ family~\cite{Kunes2010,Jang2021,Park2024,Moon2015,Day-Roberts2023}. An early DFT+DMFT study~\cite{Kunes2010} highlighted the role of chalcogen dimer states in the MIT of the NiS$_{2-x}$Se$_x$ family, placing the S-S antibonding dimer states below the Ni $3d$ UHB, suggesting that the lowest unoccupied states have predominantly S dimer character. A recent bulk-sensitive x-ray spectroscopic study~\cite{Fujinuma2024}, however, revealed discrepancies between the measured valence-band spectra and the DFT+DMFT result. A small or even negative CT energy has been inferred from Ni 2$p$ core-level photoemission and x-ray absorption studies~\cite{Laila2024,Fujinuma2024}. This suggests a substantially smaller splitting between the UHB and the top of the valence band
similar to the picture obtained in the later DFT+DMFT studies~\cite{Moon2015,Day-Roberts2023}.

We revisit the electronic structure of NiS$_2$ and NiSe$_2$ using the DFT+DMFT method in combination with photoemission spectroscopy (PES).
The relative energy of the Ni $3d$ and chalcogen $p$ states, determined by the so-called double-counting correction $\mu_{\rm dc}$, 
is treated as an adjustable parameter obtained by comparison to the valence-band PES data, an approach that has recently been applied to several TM and actinide compounds~\cite{Takegami2022,Marino2024,Sundermann2025}. 
The refined DFT+DMFT model reproduces not only the valence-band spectra but also the core-level PES and x-ray absorption spectra. Based on this realistic model, we examine the low-energy electronic structure, CT excitations, and magnetic moment of NiS$_2$. Finally, we compare the electronic structures of NiS$_2$ and NiSe$_2$ and discuss the implications for the MIT in the NiS$_{2-x}$Se$_x$ family.

\section{Methods}

({\it Theory.})~We performed DFT+DMFT calculations for NiS$_2$ and NiSe$_2$ using the computational implementation described in Refs.~\onlinecite{Hariki2017,Hariki2020}. DFT calculations were carried out for the experimental crystal structures within the local density approximation using the WIEN2k package~\cite{wien2k}. A tight-binding (TB) model spanning the Ni 3$d$, S/Se $p_\pi$, and S/Se $p_\sigma$ states was constructed from the DFT bands using wien2wannier and wannier90~\cite{wien2wannier,wannier90}. To investigate the role of the S-S (Se-Se) dimers, we also considered a reduced Ni-S/Se $p_\pi$ model in which the S/Se $p_\sigma$ orbitals were excluded from the TB Hamiltonian numerically. The tight-binding Hamiltonian was supplemented with local electron-electron interactions within the Ni 3$d$ shell, using the Hubbard parameter $U=5.0$~eV and Hund's coupling $J=1.0$~eV, following Ref.~\onlinecite{Kunes2010}. The interacting lattice model was solved within DMFT. The hybridization-expansion continuous-time quantum Monte Carlo impurity solver~\cite{Werner2006,Gull2011,Boehnke2011,Hafermann2012} was employed to compute the self-energy of the Ni 3$d$ electrons from the Anderson impurity model (AIM).
All DMFT calculations were performed at a 
temperature of 580~K.
After convergence of the DMFT self-consistency cycle, the self-energy was analytically continued to the real-frequency axis using the maximum entropy method~\cite{Jarrell1996}, from which the spectral functions were calculated. 
Ni 2$p$ core-level PES and XAS spectra of NiS$_2$ were calculated using the computational method provided in Refs.~\onlinecite{Hariki2017,Hariki2020}, in which the AIM with the DMFT hybridization densities was extended to include the Ni 2$p$ core orbitals and their Coulomb interactions with the Ni 3$d$ valence electrons explicitly.
 
In the DFT+DMFT methodology, the double-counting correction $\mu_{\rm dc}$ is introduced to subtract electron-electron interaction effects already included at the DFT level~\cite{Kotliar2006,Karolak2010,Haule2015}. Several schemes have been proposed in the literature, one of which was previously applied to NiS$_2$ and NiSe$_2$~\cite{Kunes2010}. We instead determine $\mu_{\rm dc}$ for the two compounds by comparing calculated DFT+DMFT spectra with experimental PES data, following a strategy that has been successfully applied to various correlated materials recently~\cite{Takegami2022,Takegami2025,Marino2024,Sundermann2025}. 
Details of the determination of $\mu_{\rm dc}$ are provided in Fig.~\ref{fig:sm_vpes} of Appendix~\ref{app:th}.

The $\mu_{\rm dc}$ controls the relative energy splitting between the Ni 3$d$ states and the S/Se $p$ states, and is therefore related to the CT energy. In the following, we use the parameter $\Delta_{dp}$~\cite{Higashi2021,Takegami2025} as a measure of the energy difference between the Ni 3$d$ and S/Se $p$ orbitals, defined as $\Delta_{dp}=\varepsilon^{\rm LDA}_d-\mu_{\rm dc}+8\times U_{\rm avr}-\varepsilon^{\rm LDA}_p = 37.71~(37.48)$~eV$-\mu_{\rm dc}$ for NiS$_2$ (NiSe$_2$). Here, $\varepsilon^{\rm LDA}_d$ and $\varepsilon^{\rm LDA}_p$ denote the on-site energies of the Ni $3d$ and S/Se $p$ orbitals in the LDA tight-binding Hamiltonian, respectively, and $U_{\rm avr}$ is the average Coulomb interaction.  Note that $\Delta_{dp}$ is similar but not exactly equal to the CT energy used in cluster-model analyses.

({\it Experiment.}) 
Hard X-ray Photoelectron Spectroscopy (HX-PES) measurements were performed at the Max-Planck-NSRRC HX-PES end station with MB Scientific A-1 HE analyzer, Taiwan undulator beamline BL12XU of SPring-8~\cite{takegami2019}. A photon energy of $h\nu=6.5$~keV with a resolution of around 250~meV were used. The X-ray Absorption Spectrsocopy (XAS) at the Ni~L$_{2,3}$ edges and Soft X-ray (SX) PES ($h\nu=1.2$~keV) measurements were performed at the MUSASHI endstation (BL-2A) of the Photon Factory. 
All measurements were performed at 300~K. Samples were fractured in-situ in order to expose a fresh surface before the measurements, and the cleanliness of the measured surface was verified with photoemission wide-scans to ensure the lack of contaminants or oxidation.

\begin{figure}[t]
\includegraphics[width=0.95\columnwidth]{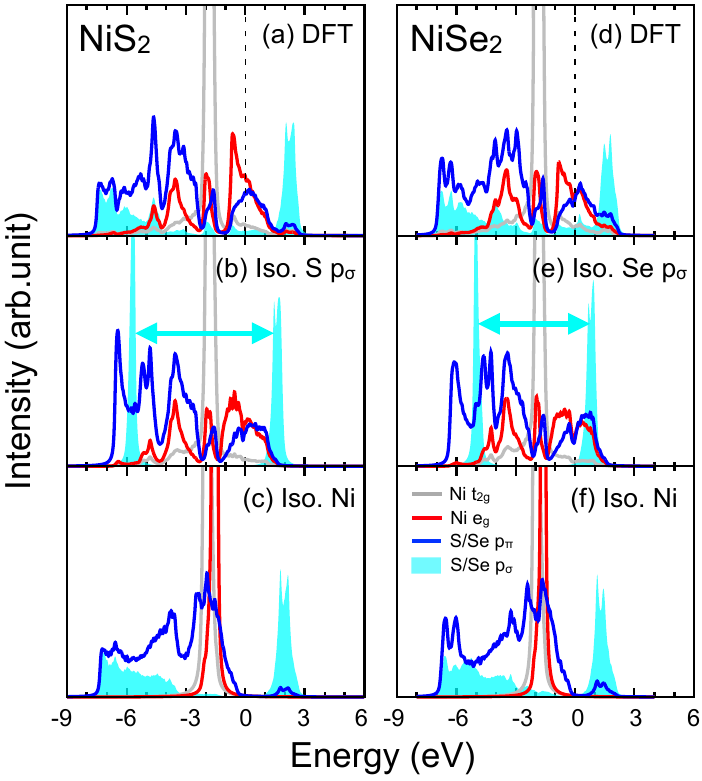}
\caption{The DFT spectral densities for NiS$_2$ (left) and NiSe$_2$ (right). The middle panels (b,e) show the results obtained after decoupling the S/Se $p_\sigma$ orbitals from the Ni 3$d$ and S/Se $p_\pi$ states. The bottom panels (c,f) show the results obtained after decoupling the Ni 3$d$ orbitals from the S/Se $p$ states (consistent with Ref.~\onlinecite{Kunes2010}).} 
\label{fig:lda}
\end{figure}


\begin{figure*}[]
\includegraphics[width=1.90\columnwidth]{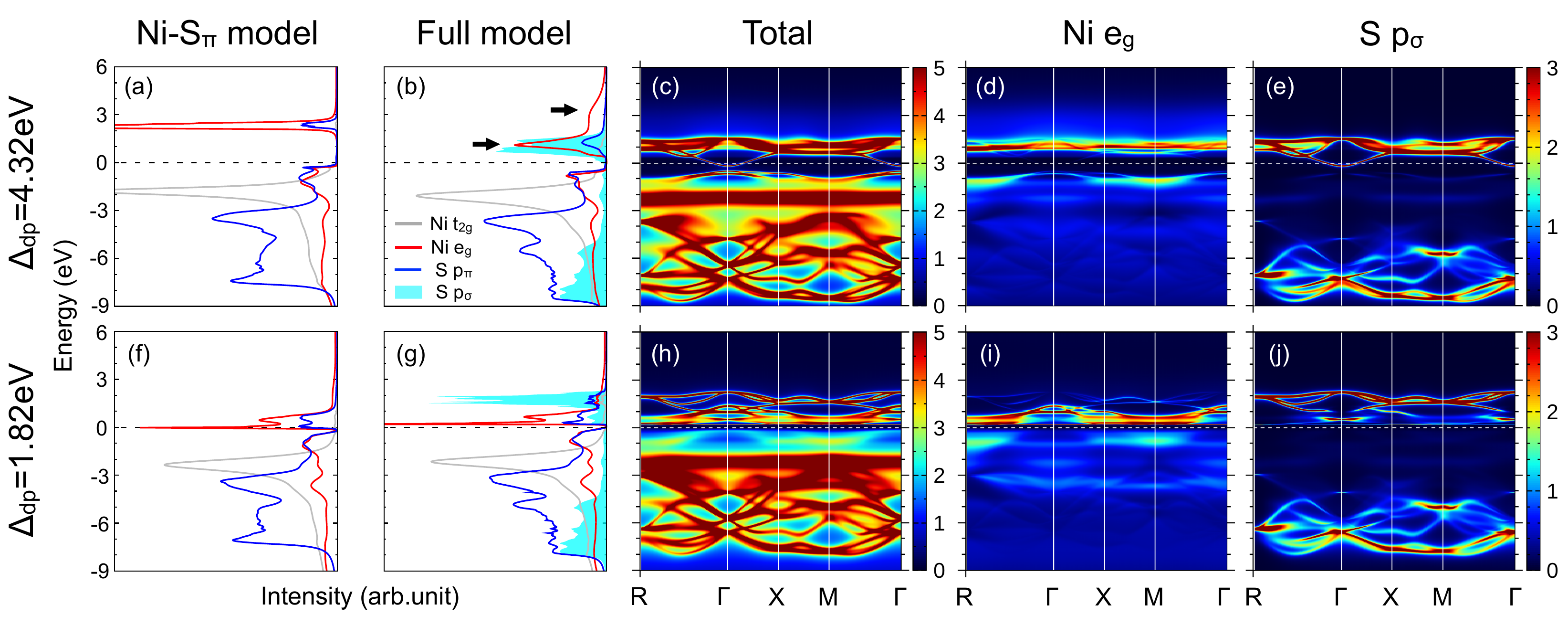}
\caption{DFT+DMFT results for NiS$_2$ obtained with $\Delta_{dp}=4.32$~eV and 1.82~eV, the latter is the optimal value. Shown are the spectral intensities for (a,f) Ni-S $p_\pi$ model, (b,g) full model, and (c,h) corresponding momentum-resolved spectra. The orbital-projected spectral intensities of (d,i) Ni $e_g$ and (e,j) S $p_\sigma$ orbitals are also shown.}
    \label{fig:akw}
\end{figure*}

\section{Results}

\subsection{DFT+DMFT for NiX$_2$ (X = S, Se)}

Figure~\ref{fig:lda}(a) shows the DFT valence-band spectra of  NiS$_2$. The S 3$p$ states are decomposed into $p_\pi$ and $p_\sigma$ orbitals perpendicular and parallel to the dimer, respectively. Thanks to the nearly right angle ($104^\circ$) between the S--S dimer and the Ni--S bond, the Ni $e_g$ orbitals hybridize primarily with the S $p_\pi$ orbitals (a $\sigma$-bond), while the Ni $t_{2g}$ orbitals hybridize more strongly with the S $p_\sigma$ orbitals (a $\pi$-bond). Figure~\ref{fig:lda}(b) presents the spectra of the model with S $p_\sigma$ decoupled~\footnote{The hopping between the S/Se $p_\sigma$ orbitals and the Ni 3$d$ and S/Se $p_\pi$ states in the wannierized DFT Hamiltonian is set to zero.} from the rest of the system, exposing the bonding-antibonding splitting on the dimer. The splitting is sufficiently large to place the antibonding state above $E_F$.
In Fig.~\ref{fig:lda}(c), where the Ni 3$d$ orbitals are isolated, the local $e_g$ and $t_{2g}$ states are identified, and the bonding S $p_\pi$ band is broadened due to hybridization with the S $p_\sigma$ band.
For NiSe$_2$, the corresponding DFT results are shown in Figs.~\ref{fig:lda}(d)–(f).
The main difference from NiS$_2$ is the weaker Se--Se dimer splitting, as pointed out in Ref.~\cite{Kunes2010} and seen in Fig.~\ref{fig:lda}(e). As a consequence, the antibonding $p_{\sigma}$ dimer state is located at lower energy than in NiS$_2$.
Upon introducing electron correlations in the Ni 3$d$ shell, the UHB develops, and its relative position with respect to the S antibonding state becomes a key factor in the excitation spectra.

In order to investigate the role of the antibonding dimer states 
and the effect of 
different double-counting corrections, we have varied 
$\Delta_{dp}$ in DFT+DMFT calculations for models with (full model) and without (Ni-S$_\pi$ model) 
the S $p_\sigma$ orbitals~\footnote{
The total electron count per unit cell in the Ni-S$_\pi$ model is reduced by two compared to the full model.}. In Fig.~\ref{fig:akw}, we show the one-particle spectral functions for two representative values of $\Delta_{dp}$. The results for the full set of $\Delta_{dp}$ values can be found in Appendix~\ref{app:th}. 
The larger $\Delta_{dp}=4.32$~eV corresponds to shallower Ni $d$ levels than the small $\Delta_{dp}=1.82$~eV. While the electron-removal (occupied states) spectra are similar for all insulating solutions, resembling the spectra of NiO~\cite{Kunes2007,Hariki2020}, the electron addition states exhibit significant differences. For the deeper $\epsilon_d$ case, $\Delta_{dp}=1.82$~eV, the S $p_\sigma$ states play a role of a passive spectator with only minor impact on the gap and the Ni $d$ spectra. We will show later that this value provides the best description of the experimental observations. Inclusion of the S $p_\sigma$ states in the model with a shallower $\epsilon_d$, $\Delta_{dp}=4.32$~eV, not only reduces the gap but qualitatively modifies the Ni $d$ spectra giving rise to two peaks marked with arrows. 

\begin{figure}[t]
     \includegraphics[width=0.99\columnwidth]{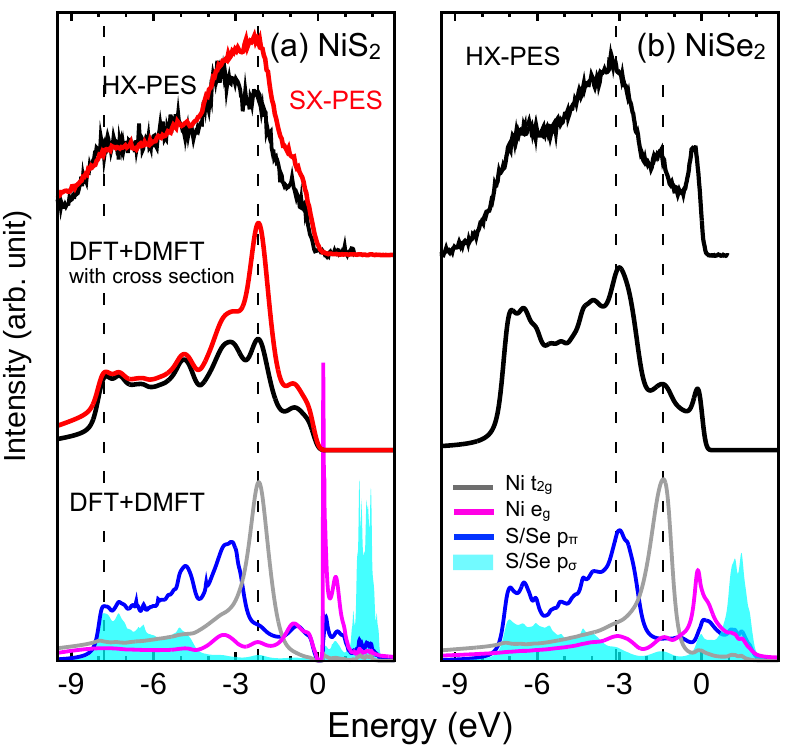}
    \caption{Comparison of the experimental PES spectra and the DFT+DMFT results for NiS$_2$ (a) and NiSe$_2$ (b). (Top) Experimental hard x-ray (HX) PES spectra (black). For NiS$_2$, the soft x-ray (SX) PES spectra (red) are also shown. (Middle) DFT+DMFT spectra obtained with the optimal $\Delta_{dp}=$1.82 (1.58)~eV, weighted by the photoionization cross sections corresponding to the experimental photon energies, multiplied by the Fermi-Dirac distribution at the experimental temperature, and broadened by the experimental energy resolution. (Bottom) Orbital-resolved DFT+DMFT spectral intensities over the full valence-band energy range.}
    \label{fig:vpes}
\end{figure}

\begin{figure}[]
     \includegraphics[width=0.90\columnwidth]{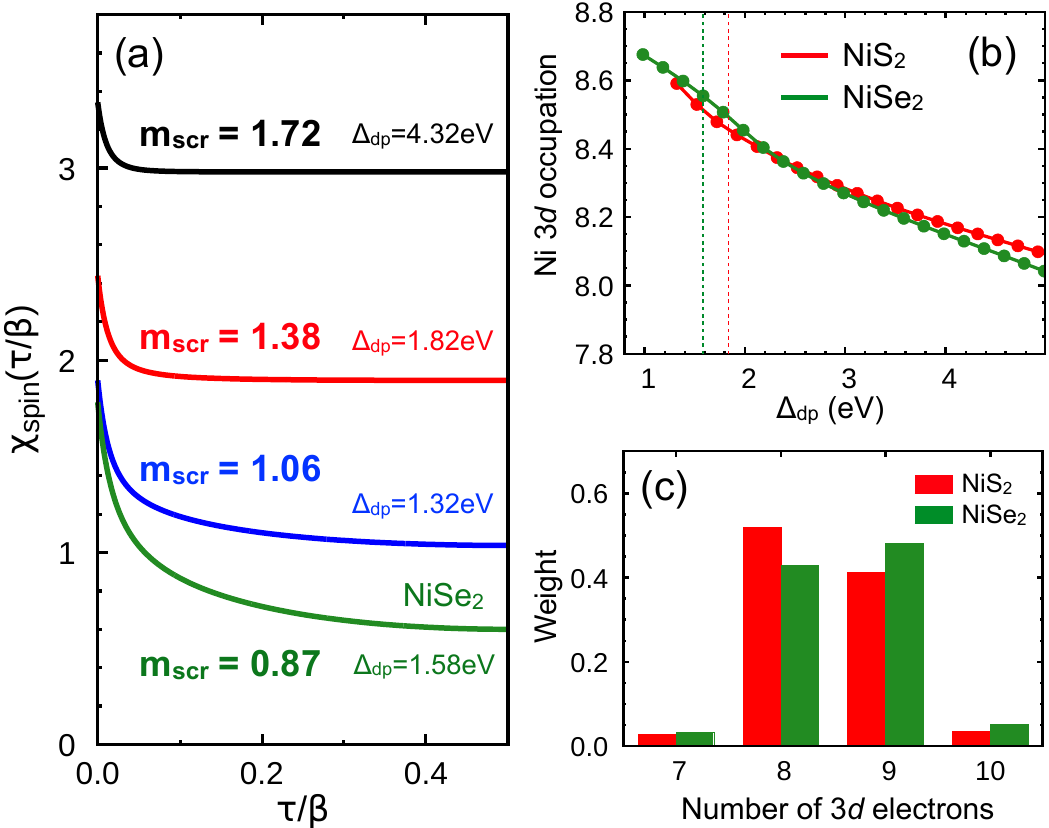}
    \caption{(a) Ni local spin-correlation function $\chi_{\rm spin}(\tau/\beta)$ in NiS$_2$, calculated using DFT+DMFT for selected values of $\Delta_{dp}$ (1.32~eV: blue, 1.82~eV: red, and 4.32~eV: black). The corresponding screened moment $m_{\rm scr}=\sqrt{\chi_{\rm spin}/\beta}$ is also indicated. The result for NiSe$_2$ at the optimal value of $\Delta_{dp}=1.58$~eV is also shown in green. Here, $\tau$ and $\beta$ denote the imaginary time and inverse temperature, respectively. (b) Ni $3d$ occupation as a function of $\Delta_{dp}$ in NiS$_2$ (red) and NiSe$_2$ (green). The optimal $\Delta_{dp}$ values are indicated by vertical dashed lines. (c) Weights of the Ni $3d^n$ configurations at the optimal $\Delta_{dp}$ values in NiS$_2$ and NiSe$_2$.}
    \label{fig:mag}
\end{figure}


It is instructive to analyze the behavior of electron addition spectra in detail, in particular, the appearance of
additional peaks for shallower double-counting (i.e. large $\Delta_{dp}$) and the absence of S $p$ spectral weight in the high energy peak (Fig.~\ref{fig:akw}(b)), in clear contrast to the case without S $p_\sigma$ 
in Fig.~\ref{fig:akw}(a). We consider an initial state $|g\rangle=\sqrt{1-|\varepsilon|^2}|t^6e^2\rangle+\varepsilon|t^6e^3\underline{p_\pi}\rangle$, 
corresponding
to six $t_{2g}$ and two $e_g$ electrons on the Ni site with a small admixture of CT state with one electron transferred from S $p_\pi$ to Ni $e_g$.
The hybridization between Ni $t_{2g}$ and S $p_\sigma$ orbital gives rise to 
the electron-addition final states $|+\rangle=\alpha|t^6e^3\rangle+\beta|t^5e^3p_\sigma\rangle$ and $|-\rangle=\beta|t^6e^3\rangle-\alpha|t^5e^3p_\sigma\rangle$ with energies $E_+$ and $E_-$.
Several observations can be made:
i) The contributions from $|g\rangle\rightarrow |\pm\rangle$ transitions to the $e_g$ spectral functions come with weights $(1-|\varepsilon|^2)|\alpha|^2$ and $(1-|\varepsilon|^2)|\beta|^2$, 
ii) The contributions to the $p_\pi$ spectra have weights $|\varepsilon\alpha|^2$ and $|\varepsilon\beta|^2$, respectively, 
iii) The coefficients $\alpha$ and $\beta$ as well as the
energy splitting $E_+-E_-$ reflect the energy difference between the $|t^6e^3\rangle$ and $|t^5e^3p_\sigma\rangle$
states. Therefore they depend only on the site energy $\epsilon_{t_{2g}}$ of the $t_{2g}$ orbitals and interaction parameters, but not on $\epsilon_{e_{g}}$.
The dependence of the electron addition spectra on $\Delta_{dp}$ is thus explained as follows.
For all values of $\Delta_{dp}$ we have $|\alpha|>|\beta|$ with $E_+<E_-$, i.e., the lower energy peak
is $t^6e^3$-like while the high energy peak is $t^5e^3p_\sigma$-like.
For shallow Ni $d$-states the $|t^6e^3\rangle$ and $|t^5e^3p_\sigma\rangle$ states are close to resonant
and thus both peaks have non-negligible weights. The $p_\pi$ contribution is concentrated in 
the low-energy peak since $|\alpha|>|\beta|$. Shifting the Ni $d$-states deeper increases the energy
splitting between the $|t^6e^3\rangle$ and $|t^5e^3p_\sigma\rangle$ states, which leads to progressive
reduction of $|\beta|$. For the deep Ni $d$-states ($\Delta_{dp}=1.82$~eV) we have $|\beta|\approx 0$
and the presence of the S dimer $p_\sigma$ states  plays no role in the one-particle spectra of the Ni states.

The behavior for shallow Ni $d$-states is an interesting example of many-body hybridization. Here, the correlated $e_g$ electron is affected by the presence of the $p_\sigma$ states, even though the two are coupled neither by direct hybridization nor by direct electron-electron interaction. This mechanism, analyzed in
a simple model in Appendix~\ref{app:model}, is akin to the splitting in the spectra of completely empty or occupied orbitals such as electron removal spectra of $t_{2g}$ orbitals in NiO~\cite{Kunes2007} or
$j_{\rm eff}=3/2$ orbitals in Sr$_2$IrO$_4$~\cite{Arita2012} and more generally to the structure of heavy-electron bands in the Falicov-Kimball model~\cite{Anders2005,Stasyuk2005}. Explaining the $e_g$ spectral features in terms of wave functions, one may ask
how this physics is represented in the Green's function language and AIM. In principle, the discussed spectral 
feature may originate from the self-energy or the hybridization function. The comparison in Fig.~S2 of the Supplementary Material (SM)~\cite{sm} reveals that the appearance of the double peak in the electron-addition part of the $e_g$ spectra originates purely from the local self-energy.

\begin{figure}[]
\includegraphics[width=0.90\columnwidth]{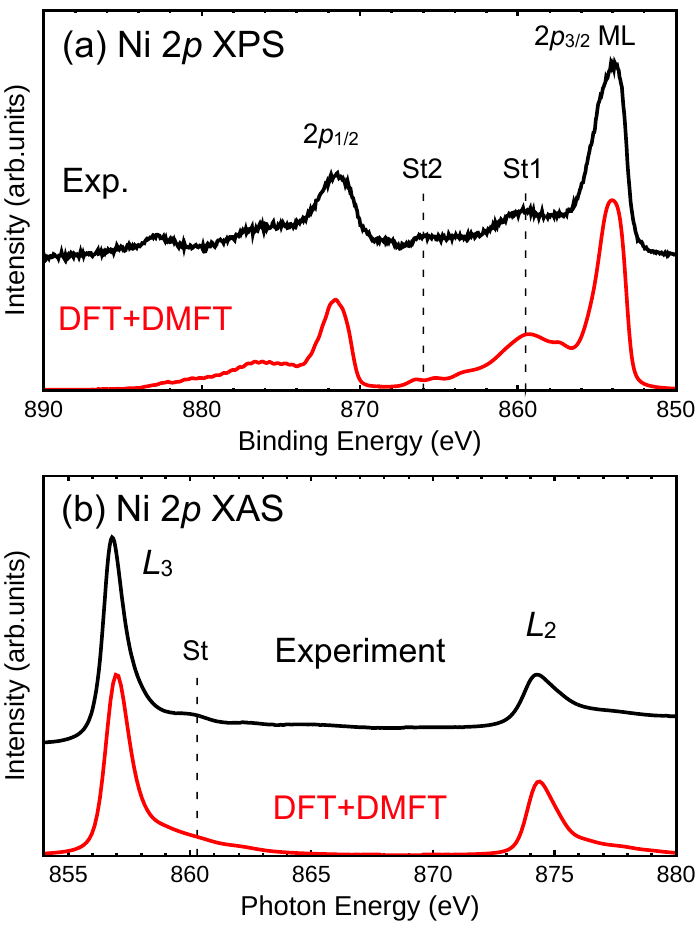}
\caption{(a) Ni $2p$ core-level PES and (b) XAS of NiS$_2$ calculated by the DFT+DMFT method using the optimal $\Delta_{dp}$ value. The experimental data are also shown for comparison.}
\label{fig:core}
\end{figure}

\subsection{NiS$_2$: Theory vs Experiment}
We determine the optimal value of $\Delta_{dp}$ to be 1.82~eV in NiS$_2$ by comparing the calculated DFT+DMFT spectra with valence-band PES spectra measured at different photon energies as shown in Fig.~\ref{fig:vpes}(a). Relative to the HX-PES spectra, the measurements in the SX range show an enhanced Ni $3d$ spectral weight with respect to that from the S $p$ due to the change in photoionization cross sections (or matrix elements)~\cite{trzhaskovskaya01,trzhaskovskaya18}. The results 
obtained with the optimal $\Delta_{dp}=1.82$~eV show good agreement with both experimental spectra. 
The different atomic-shell sensitivities of the spectra allow us to take into account not only the positions but also the orbital characters of the spectral features.
Details of the search for optimal $\Delta_{dp}$
are provided in Fig.~\ref{fig:sm_vpes} of Appendix~\ref{app:th}.

The optimized DFT+DMFT electronic structure also reproduces the Ni 2$p$ core-level PES and XAS spectra presented in Fig.~\ref{fig:core}.
In particular, the two satellite features, denoted St1 and St2 in the Ni 2$p_{3/2}$ component, are well captured.
As shown in Fig.~S5 of SM~\cite{sm}, the satellite features are reproduced only for $\Delta_{dp}$ values around the optimum, providing further support for the chosen value of $\Delta_{dp}$. Compared with cluster-model calculations~\cite{Fujinuma2024}, 
the quasi-continuous bath of AIM improves the description of the  broad spectral shape of St1
as well as the width of the main peak.

Figure~\ref{fig:mag}(b) shows the evolution of Ni 3$d$ occupancy with $\Delta_{dp}$,
which reflects the relative energies of 
$|t^6e^2\rangle$ 
and
$|t^6e^3\underline{p_\pi}\rangle$
configurations and their mixing in the ground state.
The Ni 3$d$ occupancy at the optimal $\Delta_{dp}$ is approximately 8.5, 
larger than 
8.2 in NiO~\cite{Hariki2017,Takegami2025}.
This implies a smaller CT energy in NiS$_2$ than in NiO, consistent with the recent cluster-model studies~\cite{Fujinuma2024,Laila2024}. 
Figure~\ref{fig:mag}(c) 
presents a histogram of Ni valence configurations, with a sizable contribution from the $d^9$ configuration.
The increasing 
$|t^6e^2\rangle$--$|t^6e^3\underline{p_\pi}\rangle$ mixing due to smaller CT energy leads to a reduction of the Ni local moment.
Figure~\ref{fig:mag}(a) shows the local spin--spin correlation function 
for selected values of $\Delta_{dp}$. At the optimal $\Delta_{dp}$, the screened local moment is about 1.38~$\mu_{\rm B}$, well below the 2~$\mu_{\rm B}$ expected for a Ni$^{2+}$ ion and consistent with the small ordered moment observed in the low-temperature phase~\cite{Yano2016}.

\subsection{Comparison of NiS$_2$ and NiSe$_2$}

Using the same procedure as in NiS$_2$, the optimal value of $\Delta_{dp}$ is found to be 1.58~eV for NiSe$_2$ (see Fig.~\ref{fig:sm_vpes} in Appendix~\ref{app:th}), and the resulting DFT+DMFT spectra are compared with the valence-band PES spectra in Fig.~\ref{fig:vpes}(b).
In Fig.~\ref{fig:dos_nise2_isolated_mdc} of Appendix~\ref{app:th}, we present the DMFT results for the Ni--Se$_\pi$ model (without Se~$p_\sigma$ orbitals). We do not find any appreciable difference in the Ni--S and Ni--Se covalency and the effective
bandwidth. When the dimer $p_\sigma$ orbitals are removed we observe almost identical behavior 
of the  Ni--Se$_\pi$ and Ni--S$_\pi$ models,
see Figs.~\ref{fig:dos_nise2_isolated_mdc} and \ref{fig:dos_isolated_mdc}, respectively.   
The similarity between the two materials is also preserved when the $p_\sigma$ orbitals are retained but the Se–Se dimer antibonding state is artificially shifted upward by 1 eV~ (Fig.~S6 of SM \cite{sm}).

\begin{figure*}[]
\includegraphics[width=1.90\columnwidth]{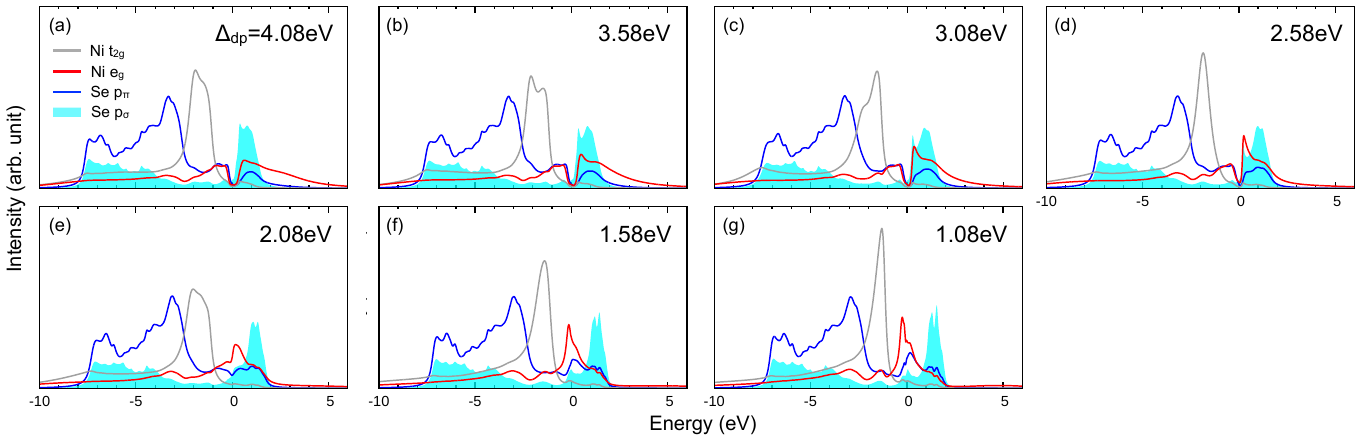}
\caption{DFT+DMFT spectra of NiSe$_2$ for selected $\Delta_{dp}$ values. The optimal value of $\Delta_{dp}$ is 1.58~eV. The $t_{2g}$ orbitals split into a singlet and a doublet in the pyrite structure, but the corresponding spectra are summed.}
    \label{fig:dos_nise2_mdc}
\end{figure*}

At their DFT positions, the Se $p_\pi$ and dimer antibonding bands overlap~\cite{Kunes2010}, rendering NiSe$_2$ a metal for all studied $\Delta_{dp}$ as can be seen from the momentum resolved spectral functions~(Fig.~S3 of SM~\cite{sm}). Nevertheless,
for shallower Ni $d$ levels $\Delta_{dp}\ge 2.58$~eV the Ni--Se$_\pi$ system develops a gap at the studied temperature.  In this case only the Se $p_\sigma$ band crosses $E_F$. The resulting internal doping of the Ni--Se$_\pi$ system
is too small to have an appreciable effect. For deeper Ni $d$ levels $\Delta_{dp}\le 2.08$~eV including the optimal $\Delta_{dp}= 1.58$~eV~\footnote{Note that this value corresponds to the same optimal $\mu_{\rm dc}$ as for NiS$_2$.} the gap between the UHB and the CT band in the Ni $d$ spectra disappears. Comparison of the spectra with and without Se $p_\sigma$ states in
Figs.~\ref{fig:dos_nise2_mdc} and \ref{fig:dos_nise2_isolated_mdc} demonstrates that the 
weak hybridization with the antibonding Se $p_\sigma$ band provides the final nudge pushing NiSe$_2$ to the metallic side. 
This mechanism is different from the pressure-induced MIT due to increase of Ni--S$_\pi$ hybridization in NiS$_2$, which pushes the valence band to lower binding energy. Shifting the UHB bottom by Se substitution or the valence band top by pressure both lead to closing of the gap and have essentially the same effect on the one-particle spectra~(Fig.~S3 of SM~\cite{sm}). 
\section{Summary}

We have studied the electronic structure of NiS$_2$ and NiSe$_2$ using the DFT+DMFT method. 
Using the experimental photoemission spectra measured with soft and hard x-rays
to determine the double-counting correction inherent to the present approach,
we have arrived at smaller charge-transfer energy than in the early work~\cite{Kunes2010},
but consistent with the later studies~\cite{Moon2015,park2026}.
The DFT+DMFT spectra accurately reproduce the characteristic satellite features in the Ni 2$p$ core-level photoemission and x-ray absorption spectra reported recently. 
NiS$_2$ and NiSe$_2$ are found to have very similar Ni--chalcogen hybridization strength, placing NiS$_2$ close to the MIT on the insulating side.
The hybridization of UHB with Se $p_\sigma$ antibonding state pushes NiSe$_2$ to the metallic side.



\begin{acknowledgements}
The authors thank K. Ozawa, T. Ohigashi, D. Shiga, and H. Kumigashira for their technical supports at BL-2A of Photon Factory. The experiment at Photon Factory was performed with the approval of KEK (Proposal No. 2024G507). This work was supported by JSPS KAKENHI Grant Numbers 25K00961, 25K07211, by the project Quantum materials for applications in sustainable technologies (QM4ST), funded as project No.~CZ.02.01.01/00/22 008/0004572 by Programme Johannes Amos Commenius, call Excellent Research, by the Ministry of Education, Youth and Sports of the Czech Republic through the e-INFRA CZ (ID:90254), by the Czech Science Foundation (Grant No. 22-22000M), and Lumina Quaeruntur fellowship LQ100102201 of the Czech Academy of Sciences. Part of the computations in this work were performed using the facilities of the Supercomputer Center, the Institute for Solid State Physics, the University of Tokyo.
\end{acknowledgements}

\begin{figure}[t]
\includegraphics[width=0.90\columnwidth]{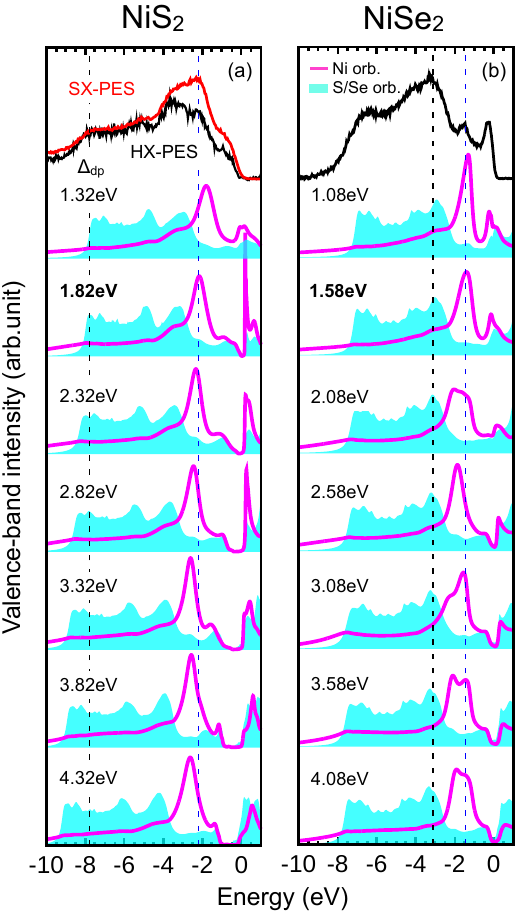}
\caption{The $\Delta_{dp}$ dependence of the DFT+DMFT spectra for (a) NiS$_2$ and (b) NiSe$_2$. Photoionization cross sections and spectral broadening are included for comparison with the experimental PES spectra (red: SX-PES data; black: HX-PES data).}
\label{fig:sm_vpes}
\end{figure}

\begin{figure*}[]
\includegraphics[width=1.90\columnwidth]{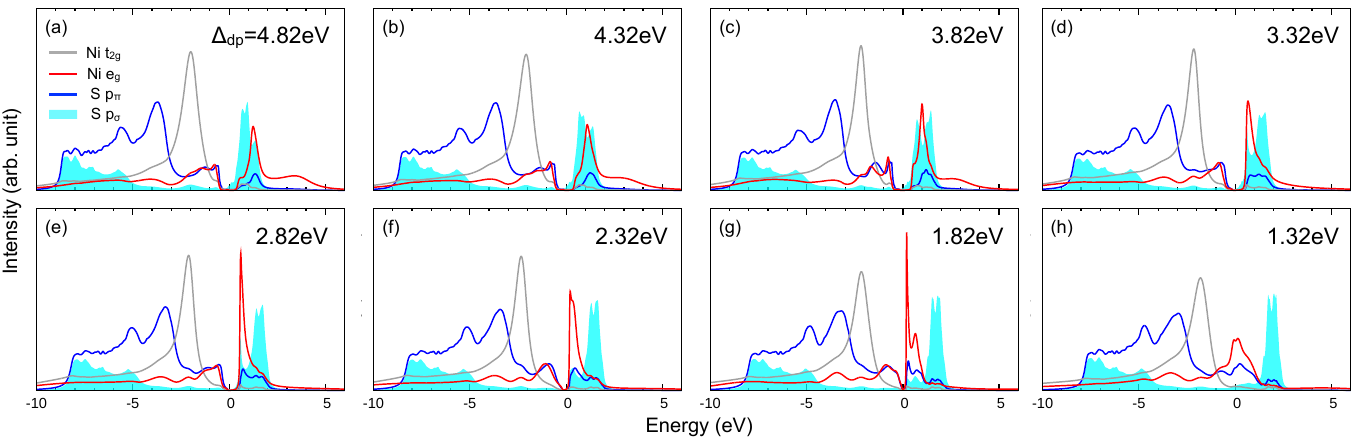}
\caption{DFT+DMFT spectra of NiS$_2$ for selected $\Delta_{dp}$ values. The optimal value of $\Delta_{dp}$ is 1.82~eV. The $t_{2g}$ orbitals split into a singlet and a doublet in the pyrite structure, but the corresponding spectra are summed.}
    \label{fig:dos_mdc}
\end{figure*}

\begin{figure*}[]
\includegraphics[width=1.90\columnwidth]{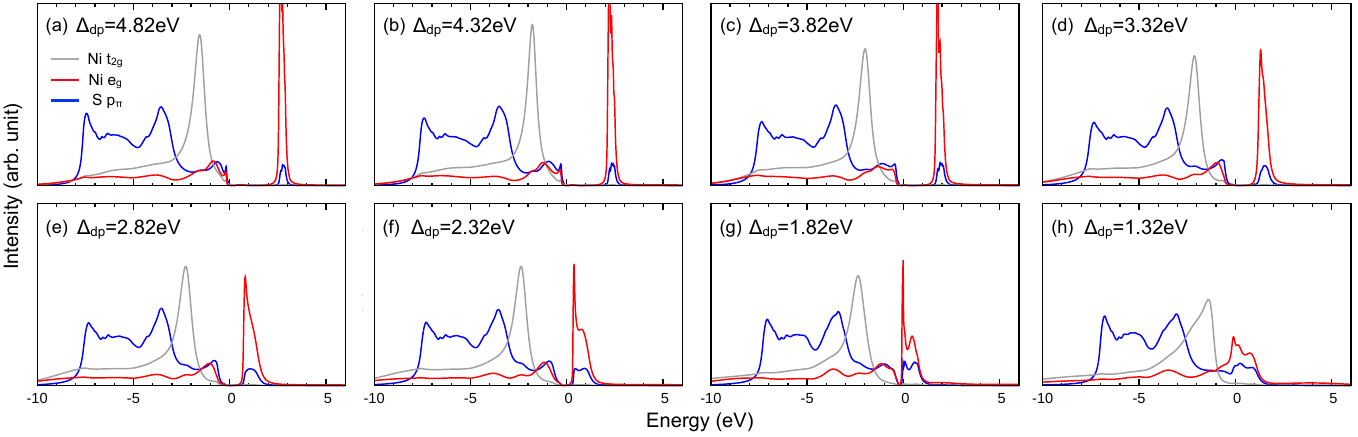}
\caption{DFT+DMFT spectra of Ni--S$_\pi$ model for selected $\Delta_{dp}$ values. The corresponding results for the full model are shown in Fig.~\ref{fig:dos_mdc}.}
    \label{fig:dos_isolated_mdc}
\end{figure*}


\begin{figure*}[]
\includegraphics[width=1.90\columnwidth]{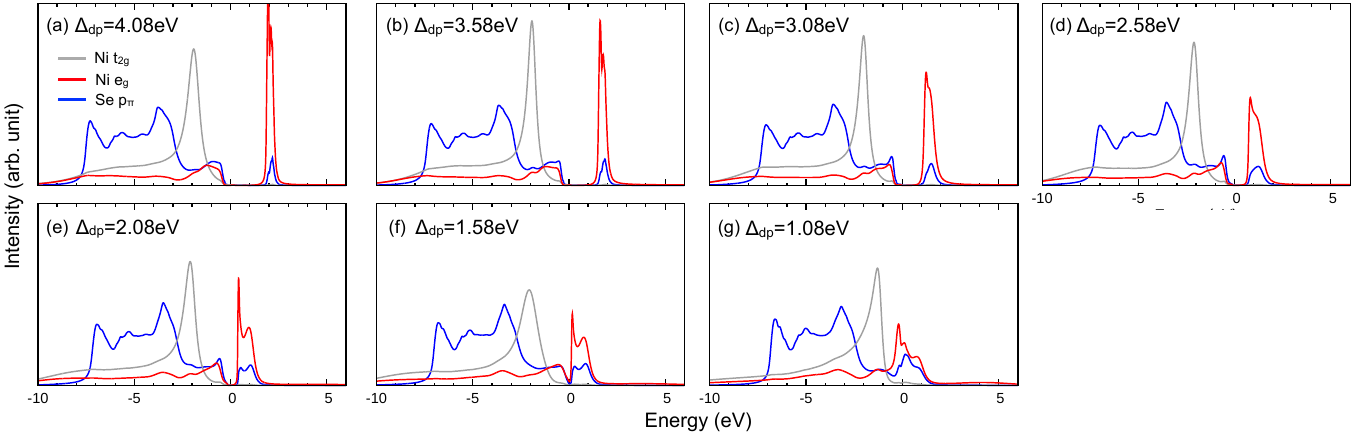}
\caption{DFT+DMFT spectra of Ni-Se$_\pi$ model of NiSe$_2$ for selected $\Delta_{dp}$ values. The corresponding results for the full model are shown in Fig.~\ref{fig:dos_nise2_mdc}.}
    \label{fig:dos_nise2_isolated_mdc}
\end{figure*}

\appendix
\section{Parameter Optimization for NiS$_2$ and NiSe$_2$}
\label{app:th}

Figure~\ref{fig:sm_vpes} shows the valence-band spectra of NiS$_2$ and NiSe$_2$ calculated using the DFT+DMFT method over a wide range of $\Delta_{dp}$ values. 
The valence-band spectra are weighted by the photoionization cross sections and broadened using the experimental resolution to facilitate comparison with the experimental data.
The adopted photoionization cross section values are shown in Table~\ref{tab_S1}. The values were calculated using the averaged tabulated $\sigma$ and $\beta$ of their respective $d$ and $p$ subshells from the values in Refs.~\cite{trzhaskovskaya01,trzhaskovskaya18}, and interpolated to the relevant energies. We note that for the NiS$_2$ experimental spectra, the Ni~$t_{2g}$ contributions appear more suppressed than expected according to the tabulated values due to the orientation dependence effects~\cite{Takegami2022od}, as the measurements were performed on single crystalline samples with the $c$ axis parallel to the linear polarization direction.
The orbital-resolved DFT+DMFT spectra without the photoionization cross sections are shown for selected $\Delta_{dp}$ values in Figs.~\ref{fig:dos_mdc} and \ref{fig:dos_nise2_mdc} for NiS$_2$ and NiSe$_2$, respectively. Note that the $t_{2g}$ orbitals split into a singlet and a doublet in the pyrite structure, but these are summed in the figure. The DFT+DMFT results for the Ni--S$_\pi$ model of NiS$_2$ and the Ni--Se$_\pi$ model of NiSe$_2$ for the corresponding $\Delta_{dp}$ values are shown in Fig.~\ref{fig:dos_isolated_mdc} and in Fig.~\ref{fig:dos_nise2_isolated_mdc}.

Since $\Delta_{dp}$, which is a linear function of the double-counting correction $\mu_{\rm dc}$, controls the relative energy of the Ni $3d$ states with respect to the S/Se $p$ states, the spectral features with dominant Ni $3d$ character exhibit systematic energy shifts as $\Delta_{dp}$ is varied. The position of the Ni $3d$-dominant peak, indicated by the blue dashed line, matches the experimental data at $\Delta_{dp}=1.82$~eV for NiS$_2$ and $1.58$~eV for NiSe$_2$. At these values, the other spectral features associated with the S/Se $p$ states are also well reproduced. These values are therefore adopted throughout the present study. 

\begin{table}[t]
    \caption{Photoionization cross sections per electron $\sigma_{\rm tot}(h\nu)$ for $h\nu$\,=\,1.2 and 6.5\,keV, calculated following Eq.\,(2) in Ref.~\onlinecite{takegami2019}, using the tabulated values in Refs.~\onlinecite{trzhaskovskaya01,trzhaskovskaya18}. All values are given in kb ($10^{-25}$\,m$^2$). }
		\label{tab_S1}
		\setlength{\tabcolsep}{14pt} 
		\begin{tabular}{ c | c c }
			state & $\sigma_{\rm tot}$(1.2\,keV) & $\sigma_{\rm tot}$(6.5\,keV) \\
			\hline \hline
			Ni\,3$d$ & 3.01  & 2.65 x $10^{-3}$ \\
			S\,3$p$ &  1.17 & 2.15 x $10^{-3}$ \\
			Se\,4$p$ &  3.04 & 2.55 x $10^{-2}$ \\
		\end{tabular}
\end{table}

\begin{figure}[]
\includegraphics[width=0.80\columnwidth]{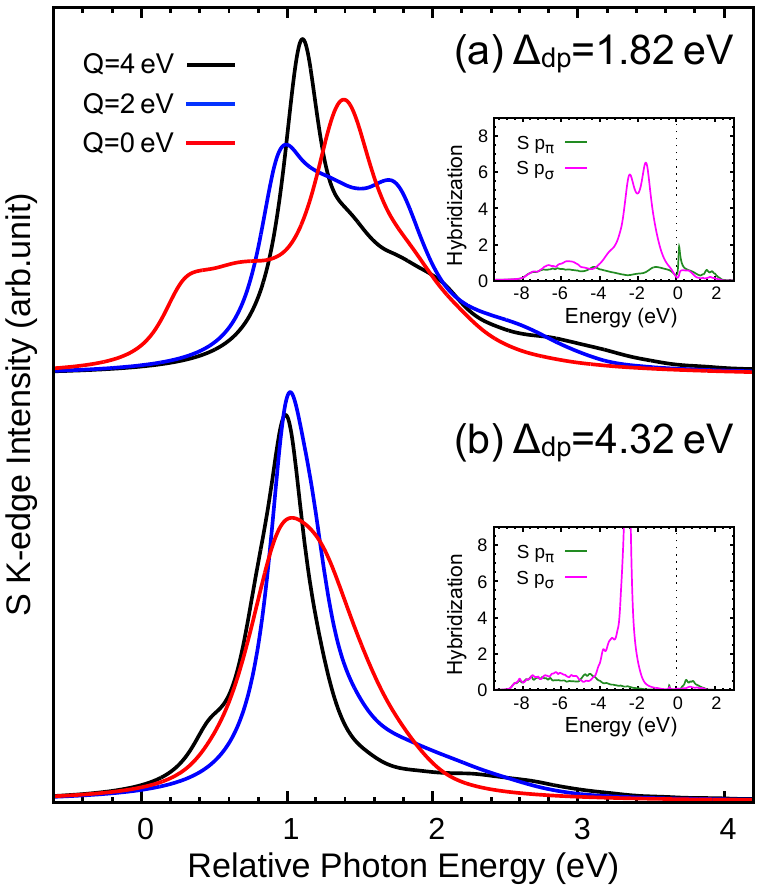}
\caption{S $K$-edge absorption spectra of NiS$_2$ simulated using the DFT+DMFT AIM method. For two selected values of $\Delta_{dp}$, (a) $\Delta_{dp}=1.82$~eV and (b) $\Delta_{dp}=4.32$~eV, the spectra are shown for different values of the core-hole potential $Q$ between the S $1s$ core-level and the S $3p$ electrons. The Lorentzian broadening with HWHM of 0.16~eV was applied to the calculated spectra. The DMFT hybridization functions for the S $p_\sigma$ (pink) and $p_\pi$ (green) states are shown in the insets.}
    \label{fig:s_k_xas}
\end{figure}

In Ref.~\onlinecite{Kunes2010}, the experimental S $K$-edge absorption spectrum of NiS$_2$ was reported. The spectrum exhibits a sharp absorption edge followed by a broader tail-like feature on the higher-energy side. In our optimized DFT+DMFT model for NiS$_2$, the unoccupied states exhibit two well-separated components with predominantly S $p_\pi$ and $p_\sigma$ character, respectively, with the latter having large spectral weight on the higher-energy side (Fig.~\ref{fig:akw}(g)), in contrast to the experimental observation. Although several factors may contribute to the relative intensities of the S $K$-edge absorption features, here we examine the possible influence of the core-hole potential between the S $1s$ core hole created by the x-ray excitation and the S $3p$ electrons in the final state of the XAS process. In transition-metal oxides, a sizable core-hole potential between the O $2p$ and O $1s$ states was inferred~\cite{Okada2006}, although its actual value is not known. We therefore simulated the spectra for different values of $Q$ (Fig.~\ref{fig:s_k_xas}). To this end, we constructed an AIM representing the sulfur site based on the DFT+DMFT results. The corresponding hybridization functions of the two S $p$ orbitals are shown in the inset of Fig.~\ref{fig:s_k_xas}. We evaluated the S $K$-edge absorption spectra using the same computational implementation adopted for the Ni core-level PES and XAS calculations. Upon introducing the core-hole potential $Q$, the spectrum is significantly modified. For a moderate value of $Q = 4$~eV, a sharp absorption peak with a broad high-energy tail appears, indicating that the higher-energy part of the S unoccupied spectral weight is suppressed by the core-hole potential. Another notable observation is that the XAS spectrum in the presence of the core-hole potential is sensitive to $\Delta_{dp}$.


\section{Simple Model for the Unoccupied States in NiS$_2$}
\label{app:model}
\begin{figure}[]
\includegraphics[width=0.90\columnwidth]{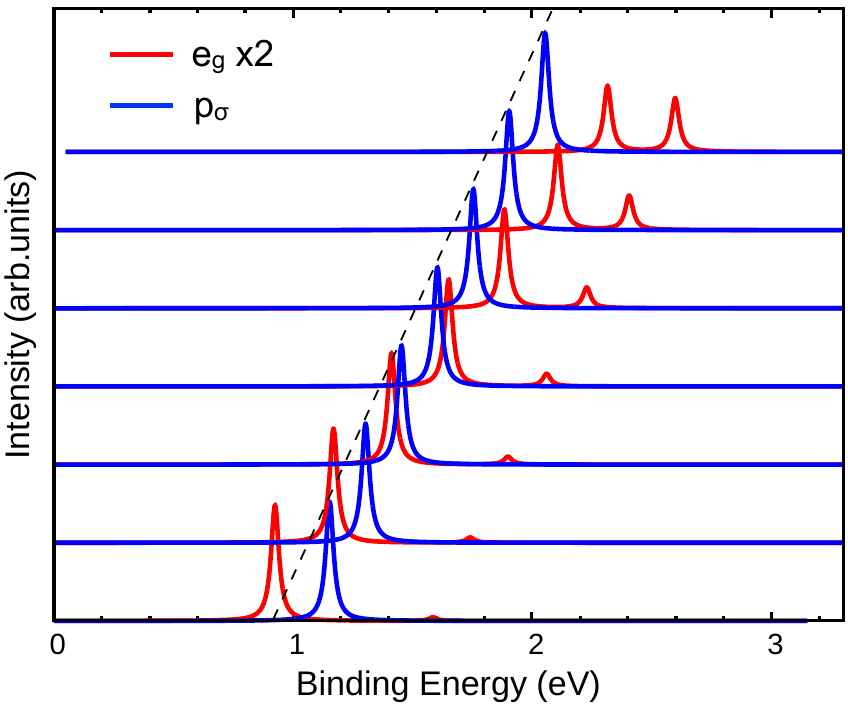}
\caption{Evolution of the spectral intensities of the $e_g$ ($b$ orbital) and $p_{\sigma}$ orbitals as a function of $\varepsilon_d$ in the simplified model.}
    \label{fig:toy}
\end{figure}

We employ a simplified model to demonstrate the impact of the antibonding S dimer states on the
electron-addition spectrum. The role of $t_{2g}$ orbitals is played by a single $a$-orbital, the $e_g$ orbitals by the $b$-orbital and antibonding 
$p_\sigma$ states by $p$-orbital. We consider a density-density interaction
and only $a$--$p$ hopping $V$ is allowed, while the $b$--$p$
hopping is set to zero. 
\begin{equation*}
\begin{split}
    H & =
 \varepsilon_d a^\dagger_\sigma a^{\phantom\dagger}_{\sigma} 
 +(\varepsilon_d+\Delta_{\text{CF}})b^\dagger_\sigma b^{\phantom\dagger}_{\sigma} 
 +\varepsilon_p p^\dagger_\sigma p^{\phantom\dagger}_{\sigma}\\
  & + \left( V a^\dagger_\sigma p^{\phantom\dagger}_{\sigma} 
  + h.c.\right)
   +U(n^a_\uparrow n^a_\downarrow+n^b_\uparrow n^b_\downarrow)\\
 & +(U-2J)n^a_{\sigma}n^b_{-\sigma}
 + (U-3J)n^a_{\sigma}n^b_\sigma,
 \end{split}
\end{equation*}
where summation over the spin index $\sigma$ is assumed. Below we use the
parameters $U=5$, $J=1$, $\varepsilon_p=1$, $\Delta_\text{CF}=0.4$, $V_a=0.1$, 
$\varepsilon_d=-9.2:-8.0$. For the parameter range of interest the ground
state is essentially $|a^2b^1\rangle$. The final states for $b$-electron addition
are spanned by $|b^2a^2\rangle$ and $|b^1a^2p^1\rangle$ forming a $2\times 2$ problem
\begin{equation}
H_{\rm FS}
=
\begin{pmatrix}
0 & V \\
V & \varepsilon_p-\varepsilon_d-3U+5J
\end{pmatrix}.
\end{equation}
Changing the double-counting in the original problem is equivalent to varying $\varepsilon_d$.
The evolution of the electron-addition part of the $b$ spectral function with $\varepsilon_d$ 
is shown in Fig.~\ref{fig:toy}. It follows the same pattern as observed for the
$e_g$ spectral function in the DMFT calculation. Clearly, the two-peak structure for the shallow 
$\varepsilon_d$ appears without direct $b$--$p$ hybridization and the splitting is controlled
by the energy of the $a$-orbitals (no dependence on $\Delta_\text{CF}$).

\bibliography{nis2}
\end{document}